\documentclass{cpbtex}
\usepackage{graphicx}
\graphicspath{{./}}
\newcommand{\secref}[1]{\hyperref[#1]{Section~\ref*{#1}\hspace{0.4cm}}}

\begin{document}

\title{Multi-dimensional kinetic modeling of laser-plasma instabilities by a Vlasov-Maxwell system\thanks{Project supported by the National Natural Science Foundation of China (Grant Nos.~12405277, U2530214, and 12475237), the Scientific Research Foundation for High-level Talents of Anhui University of Science and Technology (Grant No.~2022yjrc106), Anhui Provincial Natural Science Foundation (Grant No.~2308085QA25), the Strategic Priority Research Program of Chinese Academy of Sciences (Grant No.~XDA25050700), and the Fund of National Key Laboratory of Plasma Physics (Grant No.~6142A04230103).}}

\author{Y. Chen$^{1}$, D. J. Liu$^{2}$, X. D. Cao$^{1}$, and C. Y. Zheng$^{2,3,4}$,\\
Z. J. Liu$^{2,3}$, L. H. Cao$^{2,3,4}$, and C. Z. Xiao$^{4,5}$\thanks{Corresponding author. E-mail:~xiaocz@hnu.edu.cn}\\[2pt]
\normalsize
$^{1}$School of Electrical and Information Engineering,\\
Anhui University of Science and Technology, Huainan 232001, Anhui, China\\
$^{2}$HEDPS, Center for Applied Physics and Technology,\\
Peking University, Beijing 100871, China\\
$^{3}$Institute of Applied Physics and Computational Mathematics, Beijing 100094, China\\
$^{4}$Collaborative Innovation Center of IFSA (CICIFSA),\\
Shanghai Jiao Tong University, Shanghai 200240, China\\
$^{5}$Key Laboratory for Micro-/Nano-Optoelectronic Devices of Ministry of Education,\\
School of Physics and Electronics, Hunan University, Changsha 410082, China}

\date{}
\maketitle

\begin{abstract}
We present \texttt{PLAW-EM2d} (Two-Dimensional \textbf{PLA}sma \textbf{W}ave \textbf{E}lectro\textbf{M}agnetic Simulation), a two-dimensional Vlasov-Maxwell code for kinetic studies of laser-plasma interactions. The code combines a hybrid electrostatic-\(TE_z\) field decomposition with an Eulerian \(2D2V\) Vlasov solver for the electron distribution function and a cold closure for the out-of-plane response. Vacuum propagation, Landau damping, and a homogeneous-plasma comparison with PIC simulations benchmark the coupled solver. We then consider two representative SRS problems. In an inhomogeneous plasma, the code reproduces convective SRS backscatter with a reflectivity consistent with the Rosenbluth estimate. In a transversely extended target, it captures SRS side scattering in a self-consistent two-dimensional Vlasov simulation, resolving oblique electromagnetic and Langmuir waves that satisfy the resonance conditions. These calculations show that \texttt{PLAW-EM2d} can describe both conventional SRS backscatter and a two-dimensional side-scattering channel relevant to laser-plasma interactions in ICF.
\end{abstract}

laser-plasma interaction, side scattering


\section{Introduction}\label{introduction}
Laser-plasma instabilities are an important concern in inertial confinement fusion (ICF) and related high-energy-density plasma experiments. They can redirect laser energy, generate hot electrons, and change the spatial deposition of the driver beam \cite{ICF1,ICFn,ICF2,ICF3,D_D,I_D,H_D,S_I}. One of these instabilities is stimulated Raman scattering (SRS), in which an incident electromagnetic wave decays into a scattered light wave and an electron plasma wave \cite{kruer,Nicholson,michel,xiao,chen1,chen2}. Because SRS depends strongly on electron kinetics, it also provides a demanding test of numerical models for wave-particle interactions \cite{chapman,winjum,zhou}. In an underdense plasma, the scattered wave may propagate backward, forward, or obliquely, depending on the local matching conditions \cite{tan,zhou_side,wen,yao}. Backscatter is usually the easiest channel to diagnose, whereas side-scattered light requires a multi-dimensional description when the plasma has a finite transverse extent or the interaction region cannot be treated as quasi-one-dimensional.

Reduced envelope and coupled-mode descriptions, including three-wave simulations \cite{luo_1, chen_3} and fluid simulations \cite{lpse, lpse2}, are useful for surveying parameter trends and examining gain. They do not, however, evolve the complete velocity-space distribution function. Trapping, phase mixing, and the detailed structure of the Langmuir wave must therefore be introduced through additional modeling rather than arising directly from their basic formulation. PIC simulations provide a kinetic description and are widely used in laser-plasma studies \cite{epoch,Klimo1,Klimo2,qing,zhao_1}; finite-particle noise can nevertheless broaden weak spectral features.

Eulerian Vlasov methods provide a complementary description by evolving a continuous distribution function on a phase-space mesh. Because the representation is free of particle noise, it is well suited to spectral diagnostics and to calculations in which a weak seed must be followed over time. These methods have been used to study stimulated Raman backscattering (SRBS) \cite{liu1,wangqing1,feng1} and stimulated Brillouin backscattering (SBBS) \cite{liu2,wangqing2,feng2,deji}. One-dimensional Vlasov simulations have resolved several kinetic processes in these instabilities, including wave-particle trapping, nonlinear frequency shifts, and the mechanisms that saturate SRBS and SBBS. Their one-dimensional geometry nevertheless excludes instabilities that propagate obliquely to the pump or depend on finite transverse structure. SRS side scattering is a clear example. Both the scattered electromagnetic wave and the Langmuir wave carry a finite transverse wave number, so the interaction cannot be represented in a one-dimensional Vlasov calculation.

Fully electromagnetic two-dimensional Vlasov simulations are considerably more demanding than one-dimensional or reduced-geometry calculations. The challenge comes from the computational cost as well as the need to couple the longitudinal electrostatic response to the transverse electromagnetic branch without compromising numerical stability. Several two-dimensional Vlasov-Maxwell codes have been developed for laser-plasma simulations \cite{vm2d_1,vm2d_2,vm2d_3}. Their applications include forward Raman scattering, beatwave acceleration, relativistic transparency, filamentation, and Weibel instabilities. Those studies were not aimed specifically at ICF conditions, for which stimulated Brillouin and Raman backscattering (SBS/SRS) are among the main laser-plasma instabilities of interest.

We previously developed the two-dimensional Vlasov-Poisson code \texttt{PLAW2d} to study trapping-induced sidebands of Langmuir waves in electrostatic plasmas \cite{plaw}. Here we extend that framework to a Vlasov-Maxwell system. \texttt{PLAW-EM2d} is used for both longitudinal and transverse SRS configurations. It advances a \(2D2V\) electron distribution function together with a cold closure for the out-of-plane velocity. Its field calculation separates a longitudinal electrostatic component from a transverse \(TE_z\) electromagnetic component, retaining the electron kinetics needed for SRS while keeping two-dimensional production calculations tractable. We test the solver with vacuum propagation, standard Landau damping, and a homogeneous-plasma SRS calculation compared with PIC. We then apply it to two SRS problems. A density-gradient target is used to examine convective amplification and the associated resonant-layer picture. A separate, transversely extended target is chosen to excite side-scattered SRS. In that geometry, the scattered light and Langmuir wave both have finite transverse wave numbers, so the instability cannot be reduced to longitudinal backscatter. To our knowledge, this is the first direct demonstration of SRS side scattering in a two-dimensional Vlasov simulation. It shows that a continuum kinetic solver can describe the conventional backscatter branch and a two-dimensional Raman channel with self-consistent electromagnetic and electrostatic daughter waves.

The paper is organized as follows. \secref{methods_main_solver} describes the numerical model and the implementation of \texttt{PLAW-EM2d}, including the electrostatic--$TE_z$ field decomposition, the split Vlasov advance, source injection, and the boundary treatment. \secref{vacuum bench} reports the vacuum-propagation test of the electromagnetic branch and the Landau-damping test of the electrostatic branch. In \secref{srbs bench}, the coupled solver is compared with PIC for a uniform plasma and is then applied to convective SRS backscatter in an inhomogeneous plasma. \secref{srss bench} presents the side-scattering calculation in the transversely extended target. The main results and possible extensions of the model are summarized in \secref{conclu}.

\section{Numerical method of PLAW-EM2d}\label{methods_main_solver}

\subsection{Governing equations and hybrid field decomposition}

The \texttt{PLAW-EM2d} code solves a reduced kinetic model for laser-plasma interactions. Rather than evolving a full \(2D3V\) Vlasov-Maxwell system, it advances a two-dimensional, two-velocity (\(2D2V\)) distribution function \(f_s(x,y,v_x,v_y,t)\) for each species \(s\). A fluid-like cold velocity \(v_{z,s}(x,y,t)\) supplies the out-of-plane response. The electric field is separated into a longitudinal electrostatic part, \(\mathbf{E}_L=(E_x,E_y)\), obtained from Poisson's equation, and a transverse electromagnetic part, \((E_z,B_x,B_y)\), advanced by a \(TE_z\) Maxwell solver.

The Vlasov equation for \(f_s\) is written as
\begin{equation}
\begin{split}
\frac{\partial f_s}{\partial t}
+ v_x \frac{\partial f_s}{\partial x}
+ v_y \frac{\partial f_s}{\partial y}
+ \frac{q_s}{m_s}\left(E_x-v_{z,s}B_y\right)\frac{\partial f_s}{\partial v_x} \\
+ \frac{q_s}{m_s}\left(E_y+v_{z,s}B_x\right)\frac{\partial f_s}{\partial v_y}
= C_s,
\end{split}
\label{eq:main_vm_vlasov}
\end{equation}
where \(C_s\) is a Krook-type relaxation operator. Collisional effects are represented by relaxing \(f_s\) toward the local Maxwellian \(f_{s0}\) at the prescribed, spatially dependent frequency \(\nu_s(x,y)\), namely \(C_s = -\nu_s (f_s - f_{s0})\). This model includes kinetic damping and thermalization without resolving individual binary collisions. The out-of-plane velocity \(v_{z,s}\) is advanced through the cold closure
\begin{equation}
\frac{\partial v_{z,s}}{\partial t}=\frac{q_s}{m_s}E_z.
\label{eq:main_vm_pz}
\end{equation}
The charge and current moments are computed as
\begin{align}
\rho_s &= \int\!\!\int f_s\,dv_x\,dv_y, \\
J_x &= \sum_s q_s\int\!\!\int v_x f_s\,dv_x\,dv_y, \\
J_y &= \sum_s q_s\int\!\!\int v_y f_s\,dv_x\,dv_y, \\
J_z &= \sum_s q_s \rho_s v_{z,s}.
\label{eq:main_vm_moments}
\end{align}

The electrostatic and electromagnetic branches are coupled in reduced form. The net charge density \(\rho_c=\sum_s q_s\rho_s\) drives the Poisson equation, whereas the Maxwell solver is driven only by the out-of-plane current \(J_z\). The in-plane currents \(J_x\) and \(J_y\) are not returned to the electromagnetic update. This approximation still retains their compressive plasma response through the charge-density perturbation \(\rho_c\) in the electrostatic branch. At the same time, the laser-like electromagnetic branch is polarized mainly along \(z\) and is driven primarily by \(J_z\). Under the weakly relativistic, quasi-one-dimensional conditions used in this work, the omitted feedback from \(J_x\) and \(J_y\) to the Maxwell solver is treated as a higher-order correction.

The electrostatic field is obtained from Poisson's equation,
\begin{equation}
\nabla^2 \phi=-\rho_c,
\qquad
E_x=-\frac{\partial \phi}{\partial x},
\qquad
E_y=-\frac{\partial \phi}{\partial y},
\label{eq:main_vm_poisson}
\end{equation}
while the transverse \(TE_z\) fields evolve according to
\begin{align}
\frac{\partial E_z}{\partial t} &= \left(\frac{\partial B_y}{\partial x}-\frac{\partial B_x}{\partial y}\right)-J_z^{\mathrm{tot}}-\sigma(x)E_z, \\
\frac{\partial B_x}{\partial t} &= -\frac{\partial E_z}{\partial y}-\sigma(x)B_x, \\
\frac{\partial B_y}{\partial t} &= \frac{\partial E_z}{\partial x}-\sigma(x)B_y,
\label{eq:main_vm_maxwell}
\end{align}
Here \(J_z^{\mathrm{tot}}=J_z+J_z^{\mathrm{src}}\). The quantity \(J_z\) is the cold-polarization current of the plasma, and \(J_z^{\mathrm{src}}\) is a prescribed external current used to inject the pump and seed lasers. The coefficient \(\sigma(x)\) defines a numerical sponge that absorbs outgoing waves near the \(x\) boundaries. Equations~\eqref{eq:main_vm_poisson} and \eqref{eq:main_vm_maxwell} therefore form a hybrid electrostatic-\(TE_z\) field model for kinetic simulations of laser-plasma instabilities.

\begin{figure}[t]
\centering
\includegraphics[width=0.5\textwidth]{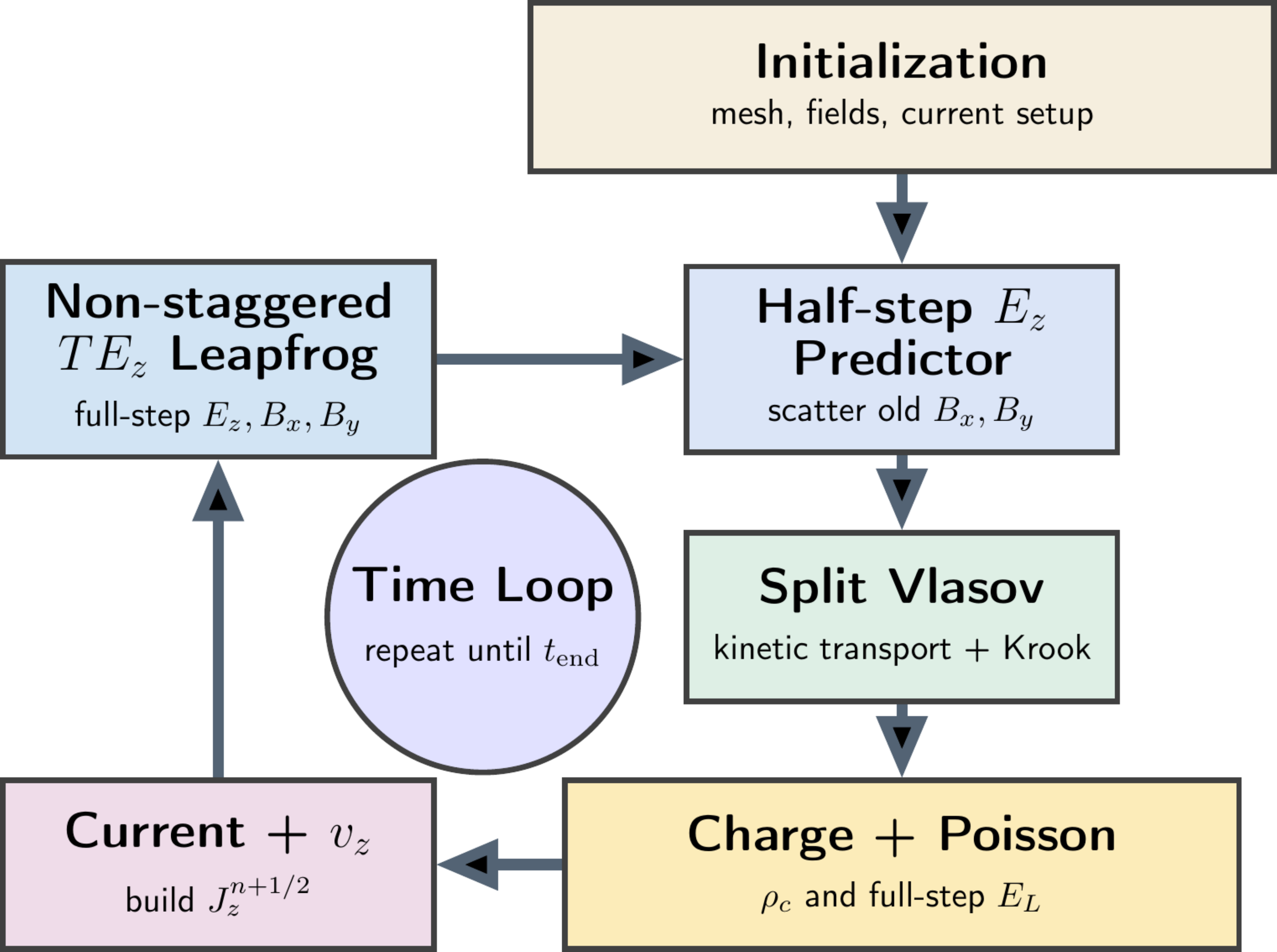}
\caption{\label{fig:main_solver_flowchart} Workflow of \texttt{PLAW-EM2d}. Initialization is performed once, whereas the closed loop is repeated during the time advance. The diagram is schematic. In the executable, the Vlasov push uses the full-step Poisson field retained from the preceding cycle, and only \(E_z\) is predicted at the midpoint before the kinetic update.}
\end{figure}

\subsection{Split Vlasov advance and cold-\texorpdfstring{$v_z$}{vz} coupling}

The code is parallelized by decomposing the physical plane across MPI ranks. Each rank stores the complete velocity mesh for its local spatial cells, and a designated root rank carries out the global field solves.

The Vlasov solver uses a directionally split sequence of one-dimensional conservative advections. A full time step of duration \(\Delta t\) is applied in the order
\begin{equation}
\mathcal{V}_x^{1/2}
\rightarrow
\mathcal{V}_y^{1}
\rightarrow
\mathcal{V}_x^{1/2}
\rightarrow
\mathcal{Y}^{1/2}
\rightarrow
\mathcal{X}^{1}
\rightarrow
\mathcal{Y}^{1/2}
\rightarrow
\mathcal{C}^{1},
\label{eq:main_vm_split}
\end{equation}
where \(\mathcal{V}_x\) and \(\mathcal{V}_y\) denote the velocity-space advections, \(\mathcal{X}\) and \(\mathcal{Y}\) denote the real-space advections, and \(\mathcal{C}\) is the Krook relaxation step. The equations solved in these substeps are
\begin{align}
\mathcal{V}_x:&\quad
\frac{\partial f_s}{\partial t}
+ a_{x,s}\frac{\partial f_s}{\partial v_x}=0, &
\mathcal{V}_y:&\quad
\frac{\partial f_s}{\partial t}
+ a_{y,s}\frac{\partial f_s}{\partial v_y}=0, \\
\mathcal{X}:&\quad
\frac{\partial f_s}{\partial t}
+ v_x\frac{\partial f_s}{\partial x}=0, &
\mathcal{Y}:&\quad
\frac{\partial f_s}{\partial t}
+ v_y\frac{\partial f_s}{\partial y}=0,
\end{align}
where \(a_{x,s}\) and \(a_{y,s}\) are the two components of the Lorentz acceleration. During the real-space advections, the velocities \(v_x\) and \(v_y\) remain fixed at the coordinates of each phase-space cell.

The field components enter the Vlasov update with a staggered time ordering. The electrostatic field is taken from the previous full step, while \(E_z\) alone is predicted at the half step before the kinetic push. The magnetic fields \(B_x\) and \(B_y\) remain at the old time level during this stage. The same predicted field \(E_z^{n+1/2}\) is used to advance the cold out-of-plane velocity \(v_{z,s}\). This splitting was chosen to maintain numerical stability without removing the kinetic coupling between the distribution and the transverse electromagnetic fields. Finally, the Krook operator is applied at the end of each time step to relax \(f_s\) toward its local Maxwellian equilibrium.

\subsection{Field update, source injection, and boundary conditions}

After the distribution function reaches time level \(n+1\), each MPI rank reconstructs the local charge density \(\rho_s^{\,n+1}\) by direct integration over velocity. The local results are assembled to form the global net charge \(\rho_c^{\,n+1}\). The root rank then obtains the electrostatic field with a two-dimensional FFT-based Poisson solver,
\begin{equation}
\hat{\phi}(\mathbf{k})=\frac{\hat{\rho}_c(\mathbf{k})}{k_x^2+k_y^2},
\qquad
\hat{E}_x=-ik_x\hat{\phi},
\qquad
\hat{E}_y=-ik_y\hat{\phi},
\label{eq:main_vm_fft_poisson}
\end{equation}
with the zero-frequency mode set to zero. The resulting full-step electrostatic field \(E_L^{\,n+1}\) is distributed to every MPI rank and retained for the next Vlasov push.

For the electromagnetic part of the update, the cold out-of-plane velocity is advanced with the same half-step field \(E_z^{\,n+1/2}\) used in the Vlasov acceleration,
\begin{equation}
v_{z,s}^{\,n+1}
=
v_{z,s}^{\,n}
+ \frac{q_s}{m_s}E_z^{\,n+1/2}\Delta t,
\label{eq:main_vm_pzfull}
\end{equation}
The new velocity is used to construct the polarization current \(J_z^{\,n+1}\). The Maxwell solver is driven by its time-centered value,
\begin{equation}
J_z^{\,n+1/2}=\frac{1}{2}\left(J_z^{\,n}+J_z^{\,n+1}\right).
\label{eq:main_vm_jzhalf}
\end{equation}
The \(TE_z\) fields are advanced with a leapfrog scheme on a non-staggered spatial grid. First, a half-step predictor is formed for \(E_z\). The electric field is then advanced through a full step using the midpoint current and the laser source, after which the magnetic field is advanced with the newly obtained \(E_z^{\,n+1}\). The pump and seed are introduced as prescribed source currents \(J_z^{\mathrm{src}}\) in the electric-field update instead of being imposed by overwriting boundary values. These currents occupy narrow strips near the left and right \(x\) boundaries and have sinusoidal time dependences with smooth spatial envelopes. A separate one-dimensional vacuum run supplies the calibration coefficient needed to launch the pump and seed at the specified amplitudes.

The boundary conditions differ between the two coordinate directions and between the kinetic and electromagnetic updates. In the transverse \(y\) direction, the Vlasov advection and electromagnetic fields can use either open or periodic boundaries. To keep the computational cost of the two-dimensional Vlasov calculations manageable, all results reported here use periodic boundaries in \(y\). In the propagation direction $x$, periodic boundaries are retained for the Vlasov solver and the Poisson equation, whereas the electromagnetic branch uses open boundaries. Outgoing waves are treated through characteristic reconstruction, with the incoming characteristic set to zero, together with the sponge \(\sigma(x)\) near both \(x\) edges. A further sponge is placed upstream of the pump source to reduce the backward-propagating branch produced there. This treatment limits artificial electromagnetic reflections while retaining an efficient periodic treatment for the kinetic and electrostatic calculations.

\subsection{Program workflow}

Figure~\ref{fig:main_solver_flowchart} summarizes the sequential operator splitting used for the full system. Each cycle from \(n\) to \(n+1\) starts with the kinetic update. The Vlasov equation uses the electrostatic field from the previous full step and a prediction of the transverse electric field at the half-integer time level. Integration of the updated distribution produces the new charge density, which supplies the global Poisson solve. After the electrostatic field has been obtained, the code advances the cold out-of-plane velocity and constructs the corresponding polarization current. The \(TE_z\) Maxwell solver then advances the transverse fields with the time-centered current. Diagnostics and output are performed after the field update. With this ordering, the field used for Vlasov acceleration remains consistent with the time centering of electromagnetic propagation while the individual substeps remain numerically stable.

\section{Benchmark of the electromagnetic branch and Landau damping}\label{vacuum bench}

We begin with a vacuum test of the electromagnetic solver before coupling it to the kinetic plasma response. This calculation exercises the non-staggered-grid leapfrog update, source-current injection, open-boundary reconstruction along \(x\), and the absorbing sponge profile. All plasma currents are switched off, leaving the prescribed pump current as the only source of the fields. The launched wave should therefore propagate out of the domain with little artificial reflection.

\begin{figure}[t]
\centering
\includegraphics[width=0.5\textwidth]{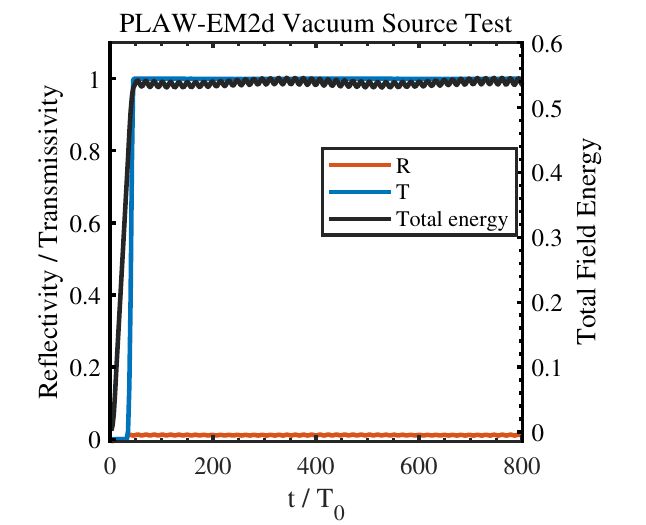}
\caption{\label{fig:vacuum_rt_test} Vacuum benchmark of the \texttt{PLAW-EM2d} electromagnetic solver. The reflectivity (corrected for the pump-only contribution via lock-in subtraction) remains small, the transmissivity approaches unity, and the total field energy remains bounded, confirming the fidelity of the source-current launch and non-staggered leapfrog Maxwell update in vacuum.}
\end{figure}

The vacuum run uses the same box-integral post-processing as the production calculations. The left box measures the left-going component, which contains the seed and backscattered light in a plasma run, while the right box measures the transmitted pump. Although the pump is launched near the left boundary, finite-amplitude leakage and numerical imperfections can leave a small pump contribution in the left-going diagnostic. We remove this contribution from the left box by lock-in subtraction. Figure~\ref{fig:vacuum_rt_test} shows that the remaining reflected signal is small and that the transmitted branch approaches unity after the source transient has crossed the domain. At the same time, the total field energy reaches a bounded plateau without secular growth. The source-current launch, open boundaries, and non-staggered leapfrog update thus reproduce the expected reflection-transmission balance before the plasma current is introduced.

\begin{figure*}[t]
\centering
\includegraphics[width=0.95\textwidth]{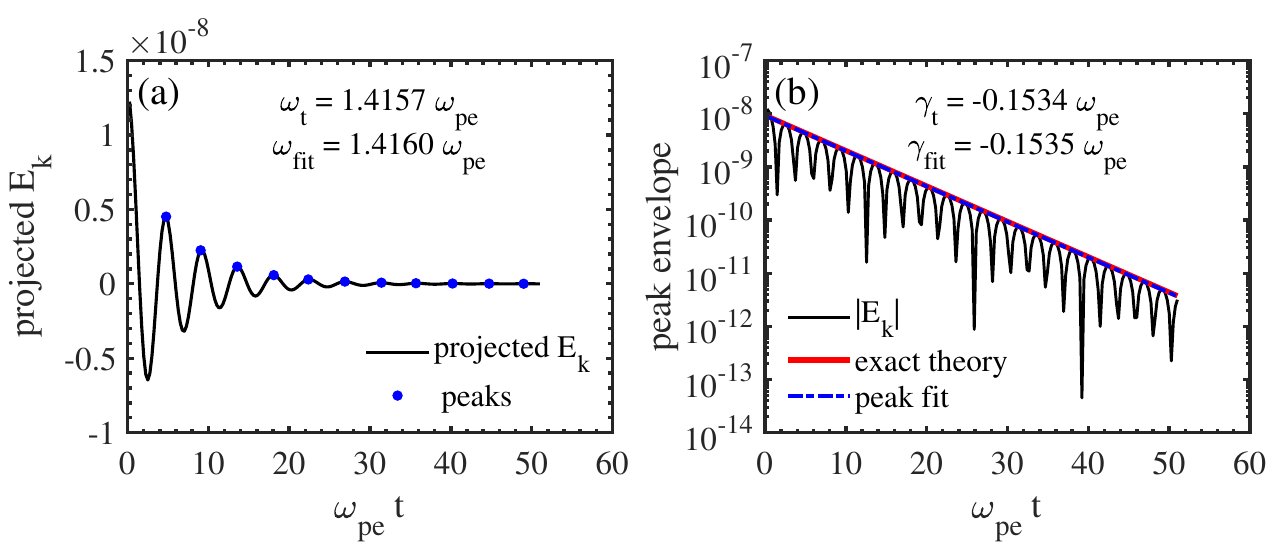}
\caption{\label{fig:landau_benchmark} Landau damping benchmark of the electrostatic branch of \texttt{PLAW-EM2d}. Left panel: projected single-mode signal \(E_{k,\mathrm{proj}}(t)\) together with the detected positive peaks used for the frequency fit. Right panel: semilogarithmic envelope of \(|E_{k,\mathrm{proj}}|\), where the exact kinetic envelope and the fitted peak envelope are plotted for comparison.}
\end{figure*}
The electrostatic solver is tested separately with the standard electron Landau-damping problem. The ions remain immobile, collisions are omitted, and the initial electron distribution is a Maxwellian with a weak single-mode density perturbation,
\begin{equation}
f_e(x,v_x,v_y,0)=\frac{n_{0e}}{2\pi v_{te}^{2}}
\exp\!\left[-\frac{v_x^{2}+v_y^{2}}{2v_{te}^{2}}\right]
\left[1+\epsilon\cos(kx)\right],
\label{eq:landau_initial}
\end{equation}
where \(\epsilon=10^{-6}\). The spatial mesh has \(N_x\times N_y=800\times160\) points, and \(N_v=128\) points are used on each half of a velocity axis. We set \(v_{te}=0.08\), \(n_{0e}=0.02553604 n_{c}\), and mode number \(m=40\), where $n_c$ is the critical density for a $351$-nm laser. These parameters give \(k\lambda_{De}=0.5\) and approximately 20 spatial cells per wavelength. The velocity domain is \(v_x,v_y\in[-0.8,0.8]=[-10v_{te},10v_{te}]\). Its discrete spacing gives a recurrence time \(T_{\mathrm{rec}}\approx1006.6\). Here, recurrence refers to the artificial rephasing that appears at late time on a discrete velocity mesh. The fitting interval ends at step 2030, corresponding to \(\omega_{pe}t\approx51.0\), and is therefore well below the recurrence time.

The raw electrostatic-field snapshots are used for the analysis. At the fixed transverse index \(j_y=81\), we project the longitudinal field onto the excited Fourier mode,
\begin{equation}
E_k(t)=\frac{1}{N_x}\sum_{j=0}^{N_x-1}E_x(x_j,y_{j_y},t)e^{-ikx_j},
\label{eq:landau_projection}
\end{equation}
and remove the initial phase over the fitting interval to obtain \(E_{k,\mathrm{proj}}(t)\). A linear fit to the times of the positive peaks gives the real frequency, while a logarithmic fit to the amplitudes of the same peaks gives the damping rate. The reference values are calculated from the exact kinetic dispersion relation,
\begin{equation}
1-\frac{\mathcal{Z}'(\zeta)}{2(k\lambda_{De})^{2}}=0,
\qquad
\frac{\omega+i\gamma}{\omega_{pe}}=\sqrt{2}\,k\lambda_{De}\zeta,
\label{eq:landau_dispersion}
\end{equation}
instead of the weak-damping approximation.

Figure~\ref{fig:landau_benchmark} shows the projected signal and its fitted envelope. For these parameters, the exact solution gives \(\omega_{\mathrm{theory}}/\omega_{pe}=1.4157\) and \(\gamma_{\mathrm{theory}}/\omega_{pe}=-0.1534\). Eleven positive peaks are resolved in the numerical signal. Fitting those peaks yields \(\omega_{\mathrm{fit}}/\omega_{pe}=1.4160\) and \(\gamma_{\mathrm{fit}}/\omega_{pe}=-0.1535\), corresponding to relative errors of \(0.0260\%\) in the real frequency and \(0.1100\%\) in the damping rate. The number of peaks follows from the duration of the run and does not indicate inadequate resolution. The frequency fit uses one positive peak from each oscillation period. Since \(2\pi/\omega_{\mathrm{fit}}\approx4.4372\,\omega_{pe}^{-1}\), the interval \(\omega_{pe}t\approx51.0195\) contains approximately \(11.5\) periods, and Landau damping suppresses the last oscillations further. The close agreement in both frequency and damping rate at \(k\lambda_{De}=0.5\) validates the electrostatic branch of \texttt{PLAW-EM2d}.

\section{Benchmark and SRS simulations in uniform and inhomogeneous plasmas}\label{srbs bench}
\begin{figure*}[t]
\centering
\includegraphics[width=0.95\textwidth]{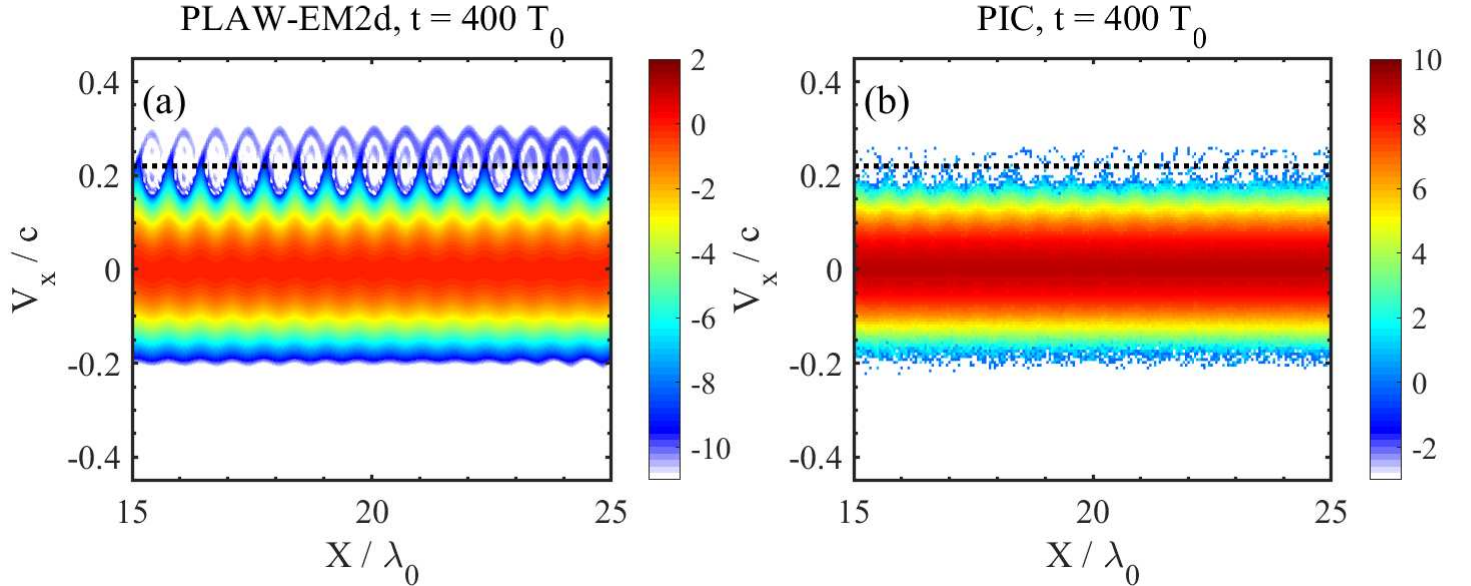}
\caption{\label{fig:vlasov_pic_phase} Benchmark comparison of the electron phase space at \(t=400\,T_0\). Left: Vlasov-Maxwell result. Right: PIC result. The dashed line marks the reference phase velocity used to identify the resonant electron response. Both simulations exhibit electron modulation around the same resonant velocity, with the distorted region confined to \(x \approx 6\lambda_0\)--\(34\lambda_0\), despite the smoother continuum solution of the Vlasov method and the noisier PIC distribution.}
\end{figure*}

\begin{figure*}[t]
\centering
\includegraphics[width=0.95\textwidth]{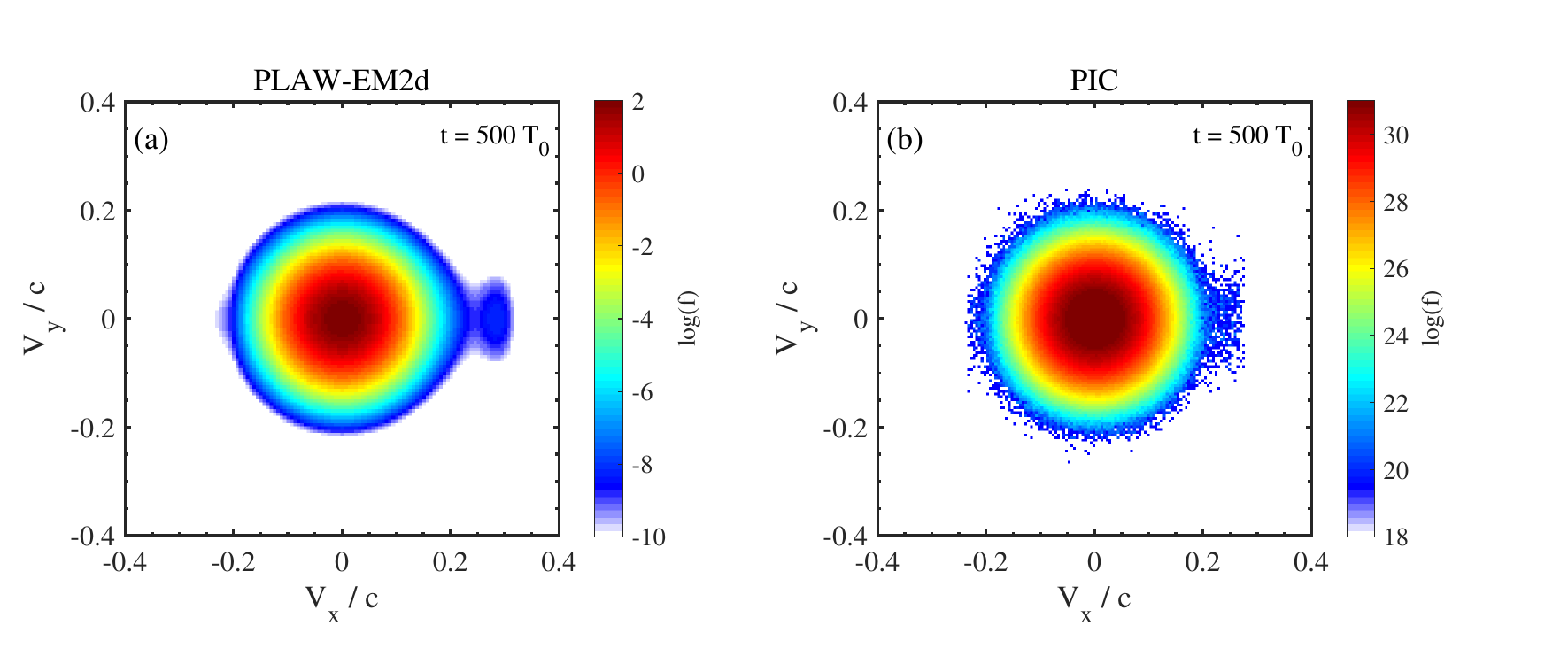}
\caption{\label{fig:vlasov_pic_vxvy} Benchmark comparison of the spatially averaged electron \(v_x\)-\(v_y\) distribution at \(t=500\,T_0\). Left: \texttt{PLAW-EM2d} result averaged over the middle plasma region. Right: PIC result. Both panels use the same velocity window, \(|v_x|,|v_y| \le 0.4c\), and are displayed in logarithmic color scale. The large-scale contour shapes and velocity spreads are consistent between the two codes, with the PIC panel showing additional granular fluctuations from finite-particle statistics.}
\end{figure*}

\subsection{Benchmark against PIC in a uniform plasma}

We next test the coupled electromagnetic-electrostatic solver against a PIC calculation under the same uniform-plasma conditions. The Vlasov method evolves a continuous distribution on a phase-space grid, whereas PIC represents the plasma with a finite ensemble of macroparticles. Agreement in the dominant structures and spectral peaks tests whether the reduced field coupling and split Vlasov advance recover the principal kinetic response.

The computational domain is \(L_x\times L_y = 40\lambda_0 \times 8\lambda_0\), resolved by \(N_x\times N_y = 800\times160\) grid points. The uniform density is \(n_e = 0.1\,n_c\), and the electron temperature is \(T_e \approx 1\ \mathrm{keV}\). Ions remain immobile, with mass ratio \(m_i/m_e = 7344\) and charge state \(Z=2\). Source currents launch a forward pump at frequency \(\omega_0\) and a counter-propagating seed at \(\omega_s = 0.6627\,\omega_0\). The physical parameters and numerical resolution are the same in the Vlasov and PIC simulations. We compare four diagnostics: the electron phase space at \(t=400\,T_0\), the spatially averaged electron \(v_x\)-\(v_y\) distribution at \(t=500\,T_0\), and the transverse field \(E_z\) and longitudinal field \(E_x\) at \(t=480\,T_0\). Together, they sample the resonant particle dynamics, the bulk velocity distribution, and both branches of the field response.

\begin{figure*}[t]
\centering
\includegraphics[width=0.95\textwidth]{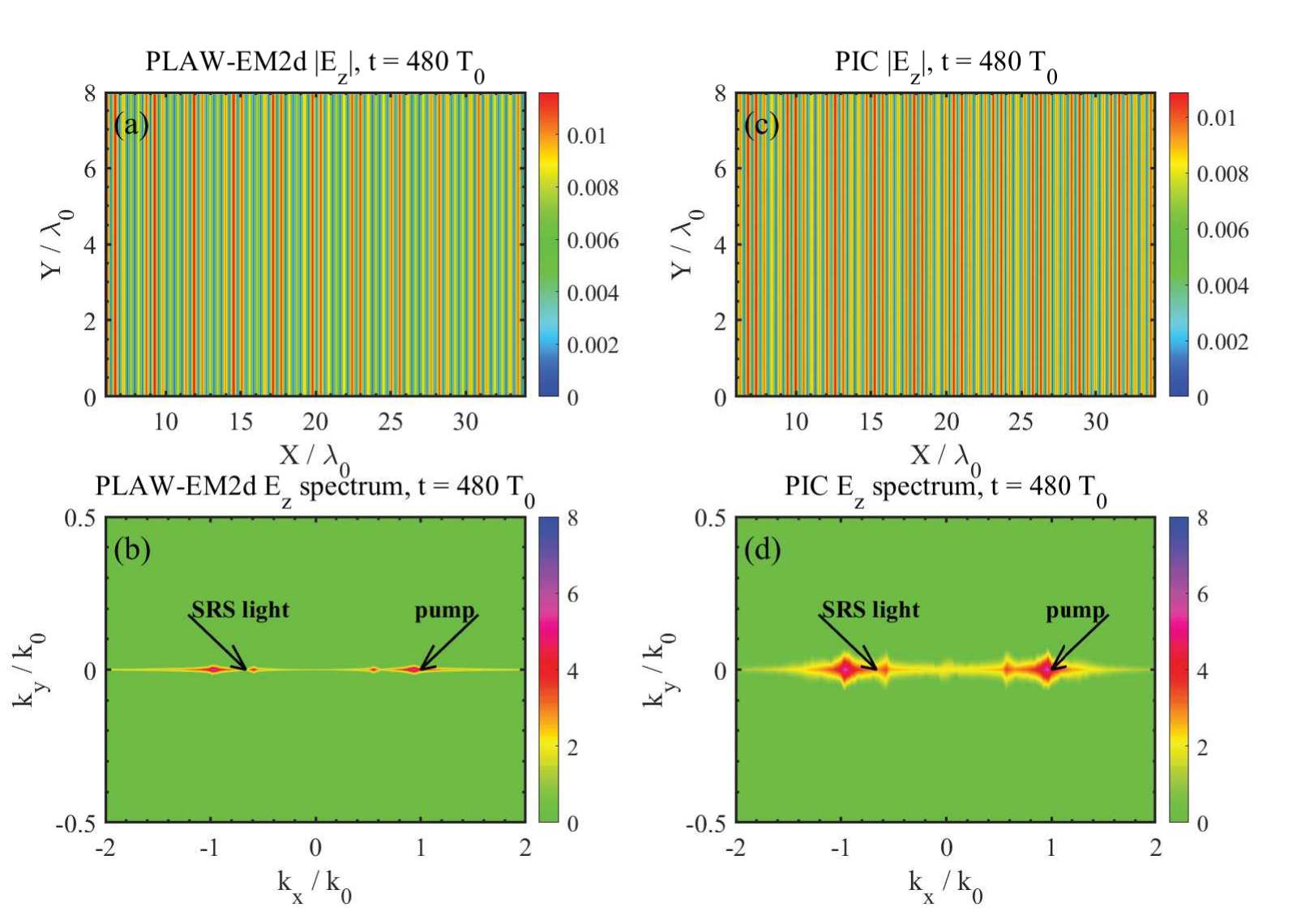}
\caption{\label{fig:vlasov_pic_ez} Benchmark comparison of the transverse electric field \(E_z\) at \(t=480\,T_0\). For each code, the upper panel shows the real-space pattern and the lower panel shows the \((k_x,k_y)\) spectrum. The dominant pump peak near \(k_x/k_0 \approx 1\) and the backscattered-light peak on the negative-\(k_x\) side appear at the same spectral locations in both calculations, although the PIC spectrum is broader due to statistical noise.}
\end{figure*}

Figure~\ref{fig:vlasov_pic_phase} gives a direct comparison of the kinetic response. At \(t=400\,T_0\), both simulations show modulation of the electron distribution around the same resonant phase velocity. In each result, the main distorted region lies within the plasma interval \(x \approx 6\lambda_0\)--\(34\lambda_0\). The Vlasov image is smoother because the continuum method evolves the distribution on a fixed phase-space grid, whereas finite macroparticle sampling produces visible granularity in the PIC image. Despite this difference in background noise, the trapped or phase-mixed electron structures have essentially the same topology. The two methods therefore identify the same interaction region and give consistent resonant electron responses even though they represent the distribution function in different ways.

Figure~\ref{fig:vlasov_pic_vxvy} compares the reduced two-dimensional velocity distributions after averaging over the central part of the plasma. At \(t=500\,T_0\), both distributions remain centered near \((v_x,v_y) = (0,0)\) and occupy the same overall velocity range. Once again, the continuum Vlasov result is smooth, while the PIC contours contain granular fluctuations from finite-particle statistics. At scales larger than this noise, the contour shape and velocity spread are consistent between the two calculations. The comparison therefore extends the agreement seen in the spatially resolved phase space to the bulk two-velocity distribution in the interaction region. It also shows no evident spurious transverse distortion associated with the reduced \(2D2V\) kinetic model.

Figure~\ref{fig:vlasov_pic_ez} compares the transverse electromagnetic response through the out-of-plane electric field \(E_z\). At \(t=480\,T_0\), both solvers produce the stripe-like interference pattern formed by the incident pump and the backscattered electromagnetic wave. Their Fourier spectra place the pump peak near \(k_x/k_0 \approx 1\) and the scattered-light peak on the negative-\(k_x\) side. Finite-particle noise makes the PIC spectrum broader and introduces additional weak spectral components. The dominant modes, however, occur at the same locations as those in the Vlasov spectrum. This agreement provides a check of the linear dispersion and the coupling between the electromagnetic branches in the \(TE_z\) solver.

\begin{figure*}[t]
\centering
\includegraphics[width=0.92\textwidth]{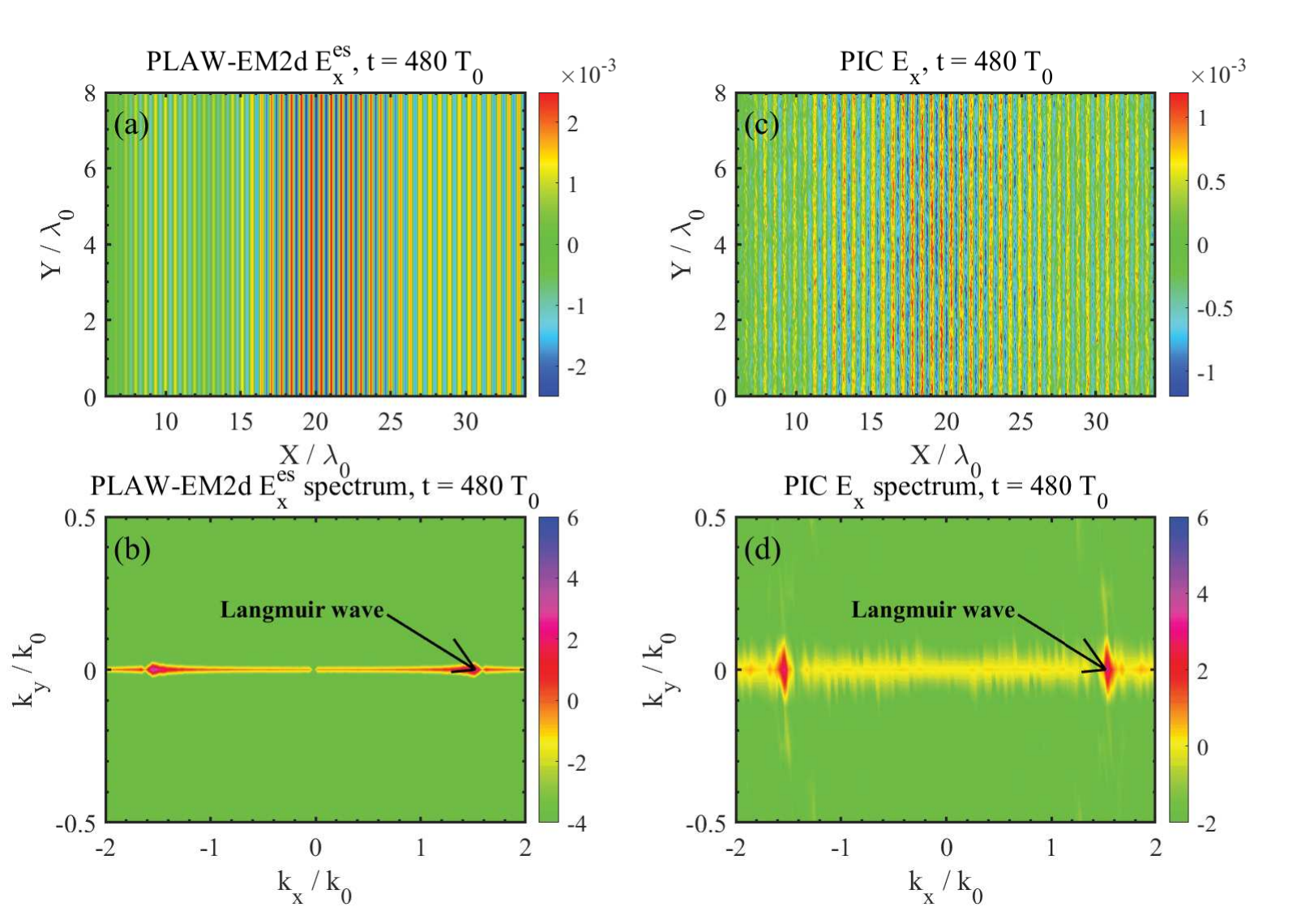}
\caption{\label{fig:vlasov_pic_ex} Benchmark comparison of the longitudinal electrostatic field \(E_x\) at \(t=480\,T_0\). For each code, the upper panel is the real-space structure and the lower panel is the \((k_x,k_y)\) spectrum. Both spectra show the same dominant Langmuir-wave peak near \(k_x/k_0 \approx 1.52\) and \(k_y \approx 0\), with the Vlasov peak being sharper and the PIC peak broader due to noise.}
\end{figure*}

The corresponding longitudinal electrostatic fields are shown in Fig.~\ref{fig:vlasov_pic_ex}. At \(t=480\,T_0\), both calculations contain a pronounced real-space pattern over the same longitudinal interval as the phase-space modulation. In the spectra, the Langmuir-wave response peaks at \(k_x/k_0 \approx 1.52\) and \(k_y \approx 0\) in both descriptions. The Vlasov peak is narrower and the PIC peak is visibly broader, but their centers coincide. This comparison is sensitive to more than the field amplitude because the Langmuir frequency and wave number depend on the electron temperature and density. The common peak position indicates that the electrostatic solver reproduces the plasma dispersion and the resonant SRS coupling condition for this case.

\begin{figure}[t]
\centering
\includegraphics[width=\columnwidth]{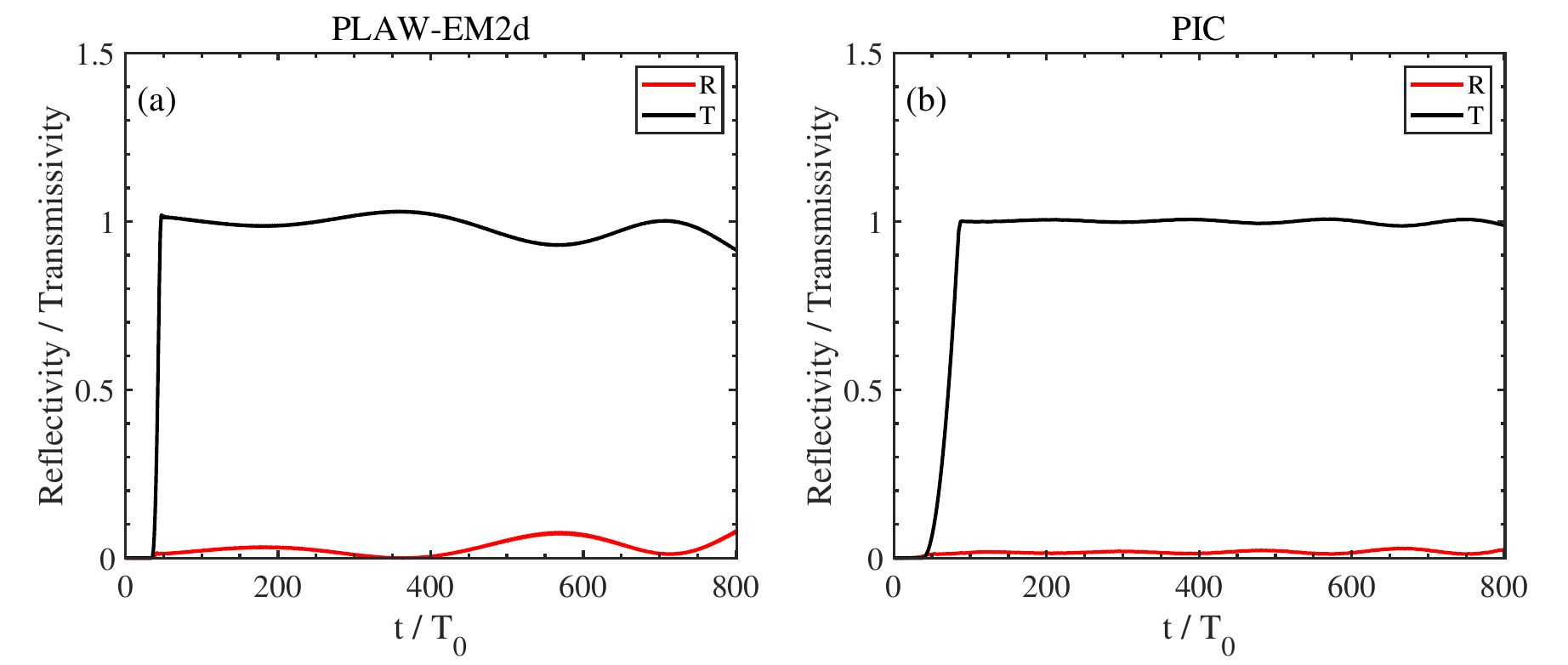}
\caption{\label{fig:vlasov_pic_ref} Benchmark comparison of the reflectivity and transmissivity histories extracted from the same box-integral diagnostics. Left: \texttt{PLAW-EM2d} result. Right: PIC result. In both codes the reflected branch remains much weaker than the transmitted branch, and the overall temporal evolution is consistent, although the PIC traces exhibit more high-frequency fluctuations.}
\end{figure}

\begin{figure*}[t]
\centering
\includegraphics[width=0.95\textwidth]{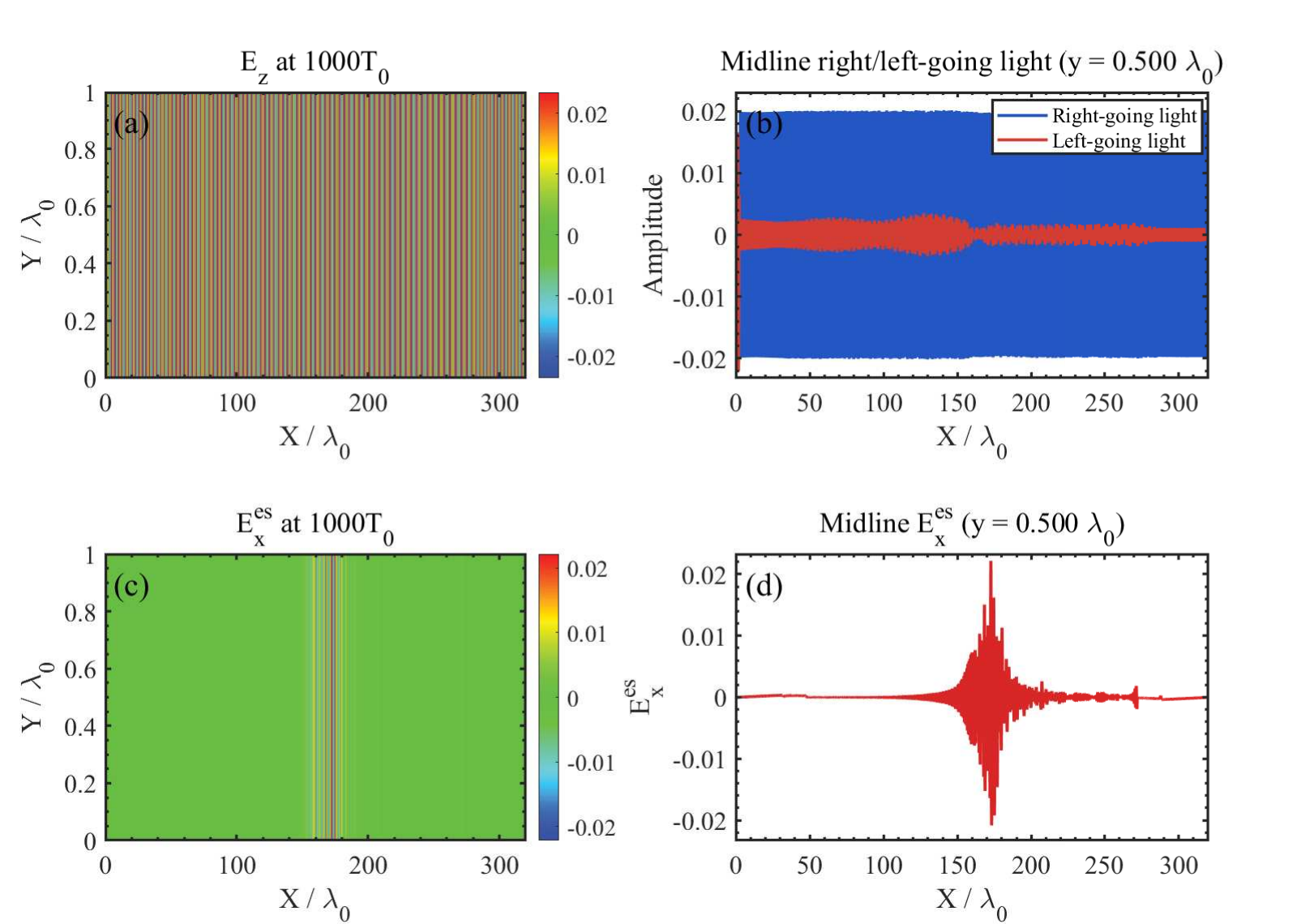}
\caption{\label{fig:inhom_fields} Field snapshot of the inhomogeneous-plasma SRS run at \(t\approx1000\,T_0\). Panel (a): transverse electromagnetic field \(E_z\), showing the interference pattern of the coexisting pump and backscattered waves. Panel (b): midline decomposition of the right-going (solid) and left-going (dashed) light components, constructed from \((E_z - B_y)/2\) and \((E_z + B_y)/2\), respectively, indicating a distinct backscattered branch. Panel (c): longitudinal electrostatic field \(E_x\), localized around the resonance region. Panel (d): midline profile of \(E_x\), showing the strongest response near the density layer where the SRS matching condition is satisfied. The electromagnetic branches extend across the entire interaction region, whereas the electrostatic response is sharply confined to the vicinity of the resonant density.}
\end{figure*}
\begin{figure}[t]
\centering
\includegraphics[width=0.65\textwidth]{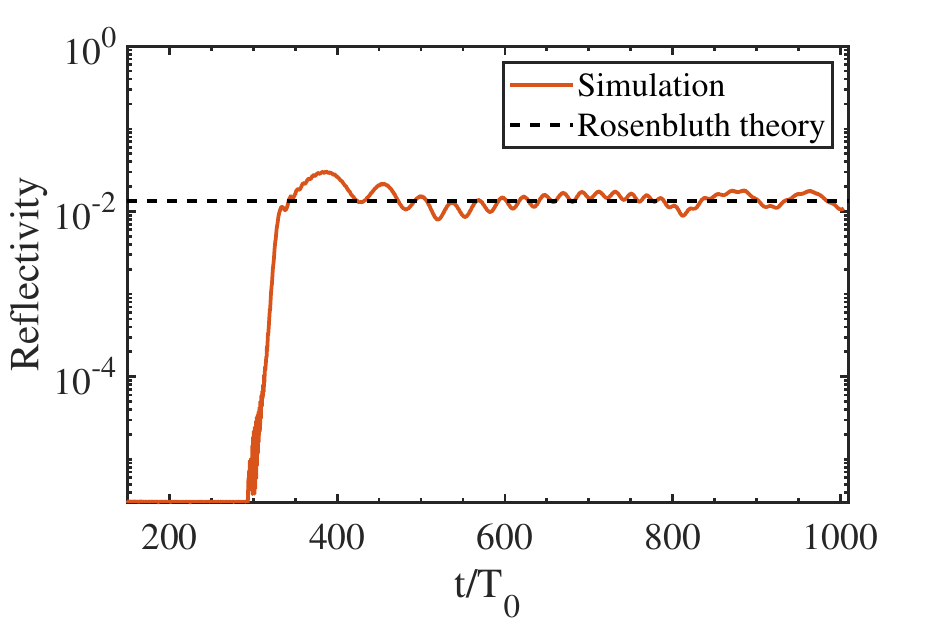}
\caption{\label{fig:inhom_ref} Time history of the backscattered reflectivity in the inhomogeneous-plasma SRS run. The solid line is extracted from the same box-integral diagnostic used elsewhere in the paper, while the dashed line is the Rosenbluth estimate evaluated from the local resonant parameters of the density-gradient case. After the initial source transient, the simulated reflectivity rises to the \(10^{-2}\) level and remains close to the theoretical expectation, confirming the convective amplification picture.}
\end{figure}
Figure~\ref{fig:vlasov_pic_ref} compares the time histories of the integrated reflected and transmitted electromagnetic branches. The PIC curves contain stronger high-frequency fluctuations because of particle noise, but both approaches give the same overall evolution. In particular, each calculation shows a weak reflected signal separated clearly from a transmitted branch near unity. The agreement is therefore not limited to instantaneous field structures and spectra; it also includes the global scattering diagnostics used in the subsequent SRS calculations.

Across Figs.~\ref{fig:vlasov_pic_phase}--\ref{fig:vlasov_pic_ref}, the two methods agree in the particle dynamics, reduced velocity distributions, transverse and longitudinal fields, and integrated scattering signals. The inhomogeneous-plasma and side-scattering calculations below therefore use the same split-field \(TE_z\) solver.

\subsection{Demonstration of SRS in an inhomogeneous plasma}

We next consider convective SRS in a nonuniform target by extending the computational domain and introducing a density gradient. The box has dimensions \(L_x\times L_y = 320\lambda_0 \times 1\lambda_0\) and is resolved by \(N_x\times N_y = 6400\times 20\) grid points. The pump and seed frequencies remain \(\omega_0\) and \(\omega_s = 0.6627\,\omega_0\), respectively. The electron temperature is again approximately \(1\ \mathrm{keV}\), and the ions are immobile. Unlike the uniform benchmark, this target has an approximately linear density profile centered at \(0.1\,n_c\), with scale length \(L_n \approx 80\ \mu\mathrm{m} \approx 228\lambda_0\). Across the main interaction region, the density rises from about \(0.051\,n_c\) on the low-density side to \(0.149\,n_c\) on the high-density side. It then tapers smoothly to vacuum near both edges.

In an inhomogeneous plasma, the SRS conditions \(\omega_0 = \omega_s + \omega_{\mathrm{LW}}\) and \(\mathbf{k}_0 = \mathbf{k}_s + \mathbf{k}_{\mathrm{LW}}\) are satisfied only near a particular density layer. At that location, the local electron plasma frequency supplies the frequency difference between the pump and scattered light. For the present parameters, the Rosenbluth estimate gives \(n_{\mathrm{res}} \approx 0.10\,n_c\), which lies near the middle of the gradient at \(x \approx 160\lambda_0\). Since the density changes along the propagation direction, the interaction is convective rather than absolute. The daughter waves grow while moving through the resonant layer and are then advected away. This behavior should produce an electrostatic response localized near resonance and an electromagnetic signal distributed over a larger part of the domain. The calculation tests whether the solver can reproduce both features and the associated amplification of the scattered light.

Figure~\ref{fig:inhom_fields} shows the fields in the inhomogeneous plasma at \(t \approx 1000\,T_0\). The transverse field \(E_z\) in panel (a) contains elongated interference stripes produced by the incident pump and the backscattered electromagnetic wave. Panel (b) separates these waves along the midline using \((E_z - B_y)/2\) for the right-going component and \((E_z + B_y)/2\) for the left-going component. The right-going pump remains stronger through most of the box. A separate left-going branch travels back toward the entrance, showing that part of the pump energy has entered the backward-scattered wave. Its amplitude builds gradually rather than remaining uniform along the target, as expected for convective amplification in a density gradient.

The electrostatic field \(E_x\) in panels (c) and (d) is localized much more strongly. Its largest amplitude occurs near the middle of the target, close to the density layer that satisfies the SRS matching condition for the selected seed frequency. Most of the \(E_x\) activity lies around \(x \approx 160\lambda_0\), with a moderate extension toward higher density, and the amplitude decreases rapidly away from this region. The density gradient explains this localization because the Langmuir wave is driven resonantly only where the local plasma frequency matches the electrostatic daughter-wave frequency. The electromagnetic pattern can propagate across the domain, while the Langmuir-wave response remains near the resonant density. Their different spatial extents are the expected field structure of convective SRS in a finite density gradient.

The global scattering diagnostic is plotted in Fig.~\ref{fig:inhom_ref}. Once the pump and seed have crossed the target and the initial source transient has subsided, the backscattered reflectivity rises from the noise level to approximately \(10^{-2}\). This level is consistent with convective amplification in the inhomogeneous target. At the late-time snapshot near \(t \approx 1000\,T_0\), the simulated reflectivity is about \(1.1\times10^{-2}\).
We compare this result with the Rosenbluth formula\cite{rosenbluth1,rosenbluth2}, which estimates the backscattered light from a steady-state convective-gain analysis in a density gradient. For SRS in an inhomogeneous plasma, the reflectivity is written as
\begin{equation} \label{wkb3}
R_{SRS} = a_{1}^{2} exp(2G_{R}),
\end{equation}
where the Rosenbluth gain is $G_{R}= \frac{\pi\gamma_{0}^{2}}{\kappa^{'}|V_{1}V_{2}|}$. Here $\kappa^{'} = \omega_{pe}^{2}/6L v_{e}^{2}k_{2}$ is the spatial wave-number detuning, $L$ is the density scale length, $k_{L}$ is the Langmuir-wave number, and $a_{1}$ is the reflected-light seed used in the simulation. The quantities $V_{1}$ and $V_{2}$ are the group velocities of the reflected light and the Langmuir wave, respectively. For the present parameters, Rosenbluth theory gives a reflectivity of about \(1.3\times10^{-2}\).

The simulation and the theoretical estimate do not agree exactly, as expected. The numerical calculation has a finite interaction length, a time-dependent source ramp, smooth tapers at the target edges, and residual temporal oscillations. These effects are absent from the ideal steady-state analysis. Even with these differences, the two reflectivities are of the same order of magnitude and follow the same convective-gain picture. This comparison, together with the localization of the Langmuir response, shows that the solver captures the expected behavior of SRS in a nonuniform target.

\begin{figure*}[t]
\centering
\includegraphics[width=0.95\textwidth]{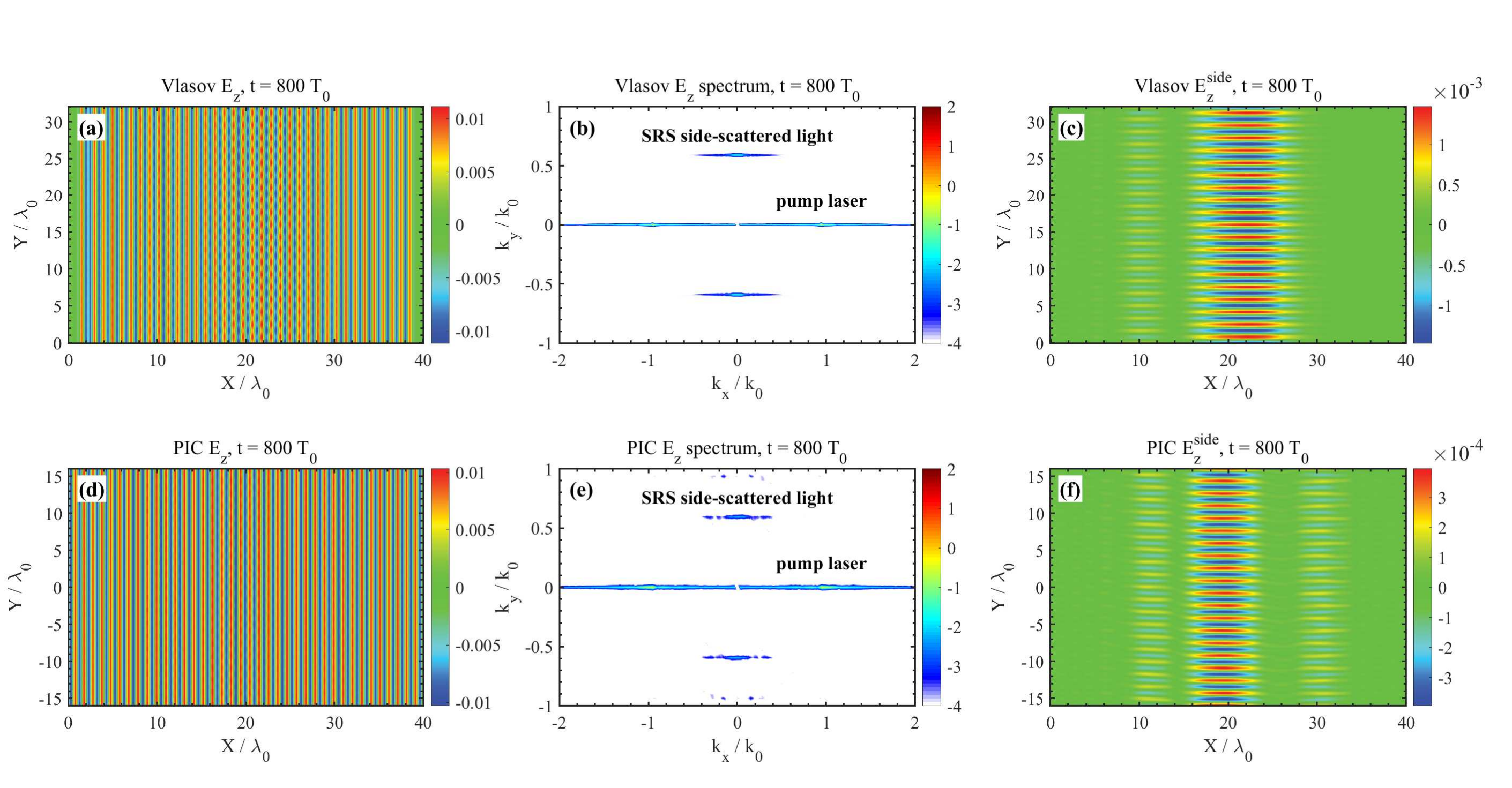}
\caption{\label{fig:side_ez} Electromagnetic diagnostics at \(t\approx800\,T_0\): \texttt{PLAW-EM2d} (top) and PIC (bottom). The \(E_z\) fields, spectra, and filtered fields show comparable oblique side-scattered packets near \((k_x,k_y)\approx(0,\pm0.593)\,k_0\); the PIC result is broader and noisier.}
\end{figure*}

\section{Demonstration of SRS side scatter in a transversely extended target}\label{srss bench}

The density-gradient calculation above tests convective backscatter in a nearly one-dimensional geometry. In an experiment, however, the laser focal spot has a finite transverse extent, and the plasma can support scattered light with a finite transverse wave number. Such light propagates at an oblique angle to the pump. This side-scattering channel requires a multi-dimensional description and is absent from one-dimensional or quasi-one-dimensional models. We therefore consider a second target with a much larger transverse width. Its parameters are chosen to excite the oblique electromagnetic and electrostatic daughter waves of SRS and to test whether \texttt{PLAW-EM2d} resolves their two-dimensional coupling.

The computational domain has dimensions \(L_x\times L_y = 40\lambda_0 \times 32\lambda_0\) and is resolved by \(N_x\times N_y = 800\times 320\) grid points. It is eight times wider in the transverse direction than the uniform-plasma benchmark. The density is centered at \(n_e = 0.1\,n_c\) and varies weakly and linearly along \(x\), with scale length \(L_n \approx 80\ \mu\mathrm{m} \approx 228\lambda_0\). Across the main plasma interval, its value changes only from \(0.094\,n_c\) to \(0.106\,n_c\). The gradient is therefore too weak to modify the local SRS matching conditions appreciably. The electron temperature is \(T_e \approx 1\ \mathrm{keV}\), and the immobile ions have \(m_i/m_e = 7344\) and \(Z=2\). We use a pump intensity of \(1.0\times10^{15}\ \mathrm{W/cm^2}\). At \(\lambda_0 = 0.351\ \mu\mathrm{m}\), the corresponding normalized vector potential is \(a_0 \approx 0.0095\), within the weakly relativistic regime.

No counter-propagating electromagnetic seed is launched in this case. We instead impose a localized electrostatic seed near the center of the domain at \((x,y) \approx (20,16)\lambda_0\). The electrostatic-field amplitude is \(10^{-3}\), and the seed has Gaussian transverse profiles. Its wave vector is \((k_x,k_y) = (0.9487, \pm 0.5933)\,k_0\). These values follow from the three-wave SRS matching conditions in a uniform plasma. For a pump of frequency \(\omega_0\) and wave number \(k_0 = \omega_0/c\) propagating along \(x\), the scattered electromagnetic wave and the Langmuir wave must satisfy
\begin{align}
\omega_0 &= \omega_s + \omega_L, \\
\mathbf{k}_0 &= \mathbf{k}_s + \mathbf{k}_L,
\end{align}
where \(\omega_s = \sqrt{k_s^2 c^2 + \omega_{pe}^2}\) is the frequency of the scattered light and \(\omega_L = \sqrt{\omega_{pe}^2 + 3k_L^2 v_{te}^2}\) is the Langmuir-wave frequency. At \(n_e = 0.1\,n_c\), the plasma frequency is \(\omega_{pe}/\omega_0 \approx 0.316\). Substitution of the selected seed vector \((k_x,k_y) = (0.9487, \pm 0.5933)\,k_0\) into \(\mathbf{k}_s = \mathbf{k}_0 - \mathbf{k}_L\) gives \((k_x,k_y) \approx (0.0513, \mp 0.5933)\,k_0\) for the scattered light. The magnitude is \(|\mathbf{k}_s| \approx 0.5955\,k_0\), which gives \(\omega_s \approx 0.674\,\omega_0\). For \(T_e \approx 1\,\mathrm{keV}\), the electron thermal speed is \(v_{te} \approx 0.044c\). The Bohm-Gross estimate then gives \(\omega_L \approx 0.328\,\omega_0\) at \(k_L \approx 1.119\,k_0\). The selected seed is consequently close to the matching condition for \(90^\circ\) side scattering, where the scattered light has only a small longitudinal component and propagates almost perpendicular to the pump.

Figure~\ref{fig:side_ez} shows the electromagnetic response of the side-scattering case at \(t \approx 800\,T_0\). The \texttt{PLAW-EM2d} results occupy the top row, and the corresponding PIC results for the same configuration are shown below. In the Vlasov real-space field [panel (a)], \(E_z\) does not have the simple longitudinal interference stripes found in the backscatter case. An oblique modulation instead develops around the seeded interaction region near the center of the domain. Its transverse spacing is \(\lambda_y \approx 1.69\lambda_0\). The Fourier spectrum in panel (b) separates this structure from the pump. Along with the dominant pump branch near \((k_x,k_y) \approx (1,0)\,k_0\), the spectrum contains sidebands at \((k_x,k_y) \approx (0, \pm 0.593)\,k_0\). These are the wave vectors expected for light scattered close to \(90^\circ\). Their values, \(|k_y|/k_0 \approx 0.593\) and \(k_x/k_0 \approx 0\), indicate propagation almost perpendicular to the pump. The widths are approximately \(\Delta k_x/k_0 \approx 0.05\) and \(\Delta k_y/k_0 \approx 0.03\), consistent with the finite size of the seeded region through \(\Delta k \sim 1/L\).

Panel (c) isolates the side-scattered field by retaining \(|k_x| \le 0.35\,k_0\) and \(0.20\,k_0 \le |k_y| \le 1.00\,k_0\), while removing the near-axis band \(|k_y| \le 0.18\,k_0\). In real space, the filtered modes form a localized electromagnetic packet centered near \((x,y) \approx (20,16)\lambda_0\). The packet extends mainly in the transverse direction, over approximately \(15\lambda_0\) in \(y\) and \(8\lambda_0\) in \(x\). Its position and orientation show that the oblique wave is emitted from the seeded interaction region and travels outward in the transverse direction. This behavior differs from pump leakage or a spatially distributed numerical background.

The PIC diagnostics in panels (d)--(f) provide an independent comparison. Although the particle calculation has a much noisier background, its real-space field also develops an oblique packet near the seed. The PIC spectrum contains the same pair of modes at \(k_y \approx \pm 0.593\,k_0\) and \(k_x \approx 0\). When the same spectral filter is applied, the PIC side-emitted component remains localized around the interaction region. The finite-\(|k_y|\) branches therefore appear in both kinetic descriptions and are not a feature peculiar to the noise-free Vlasov calculation.

\begin{figure}[!htbp]
\centering
\includegraphics[width=0.62\textwidth]{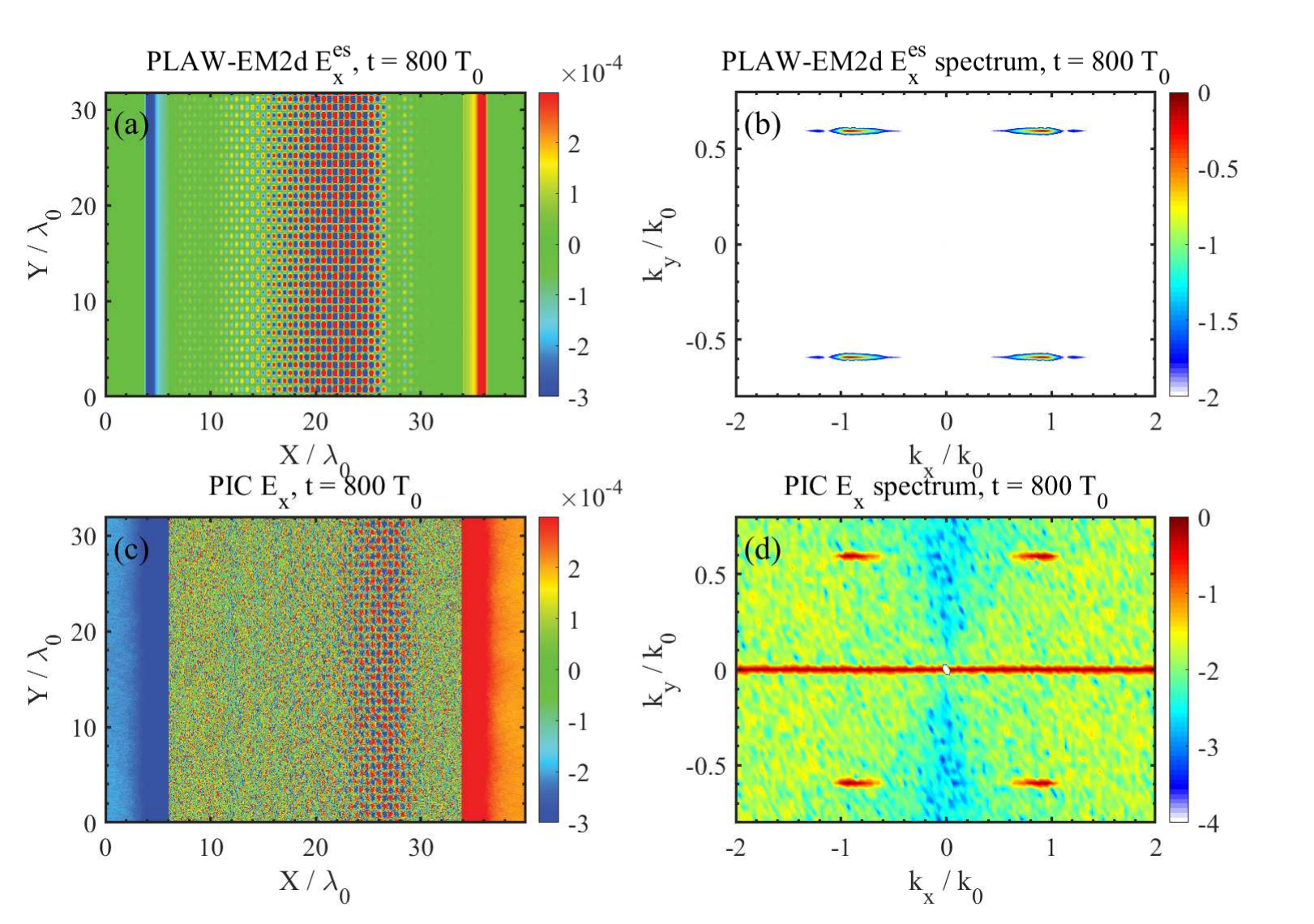}
\caption{\label{fig:side_ex} Electrostatic diagnostics at \(t\approx800\,T_0\): \texttt{PLAW-EM2d} (top) and PIC (bottom). The \(E_x\) fields and spectra show the same oblique Langmuir response near \((k_x,k_y)=(0.9487,\pm0.5933)\,k_0\), consistent with SRS matching.}
\end{figure}

The electrostatic response in Fig.~\ref{fig:side_ex} is used to determine whether the side-scattered electromagnetic modes are associated with SRS. The top row again gives the \texttt{PLAW-EM2d} results, and the corresponding PIC data are shown in the bottom row. In panel (a), the \(E_x\) field is localized in the same central region as the side-scattered \(E_z\) packet, with its strongest response close to the seeded Langmuir wave. The real-space modulation has a longitudinal wavelength \(\lambda_x \approx 2\pi/(0.9487\,k_0) \approx 1.05\lambda_0\). This wavelength agrees with the \(k_x\) component of the imposed electrostatic seed.

The Fourier spectrum in panel (b) provides a more quantitative comparison. Its dominant electrostatic peaks lie at \((k_x,k_y) = (0.9487, \pm 0.5933)\,k_0\). Their amplitudes are larger than the \(k_y=0\) component at the same \(k_x\) by factors of \(4.2\) and \(3.8\), respectively. The peak locations coincide with the wave vector of the seeded Langmuir wave, showing that the seed wave number is retained without an observable numerical shift. They also satisfy the SRS relation \(\mathbf{k}_L = \mathbf{k}_0 - \mathbf{k}_s\), where \(\mathbf{k}_s \approx (0, \pm 0.5933)\,k_0\) corresponds to the electromagnetic sidebands in Fig.~\ref{fig:side_ez}. Both the frequency and wave-vector relations between the pump, scattered light, and Langmuir wave are therefore consistent with SRS. This matching distinguishes the sidebands from a directly driven electromagnetic response or a parametric decay into a different channel.

The electrostatic peaks have widths \(\Delta k_x/k_0 \approx 0.04\) and \(\Delta k_y/k_0 \approx 0.02\). They are slightly narrower than the electromagnetic sidebands, as expected from the localized seed profile. Panel (a) also shows a small real-space asymmetry between the \(+k_y\) and \(-k_y\) branches, visible as a tilt toward the upper right. This feature suggests a residual transverse gradient that weakly breaks the symmetry. The two spectral amplitudes nevertheless remain comparable within \(\sim 10\%\).

Panels (c) and (d) show the same response in the independent PIC calculation. The PIC \(E_x\) field is concentrated in the central interaction region, and its spectrum is also dominated by oblique peaks near \((k_x,k_y) = (0.9487, \pm 0.5933)\,k_0\). These peaks are broader and lie on a noisier background than the Vlasov peaks, but their central positions agree. They also coincide with the Langmuir-wave vector predicted by the SRS matching condition. The two codes therefore identify the same oblique electrostatic daughter wave associated with the side-scattered light.

\begin{figure}[htbp]
\centering
\includegraphics[width=0.75\textwidth]{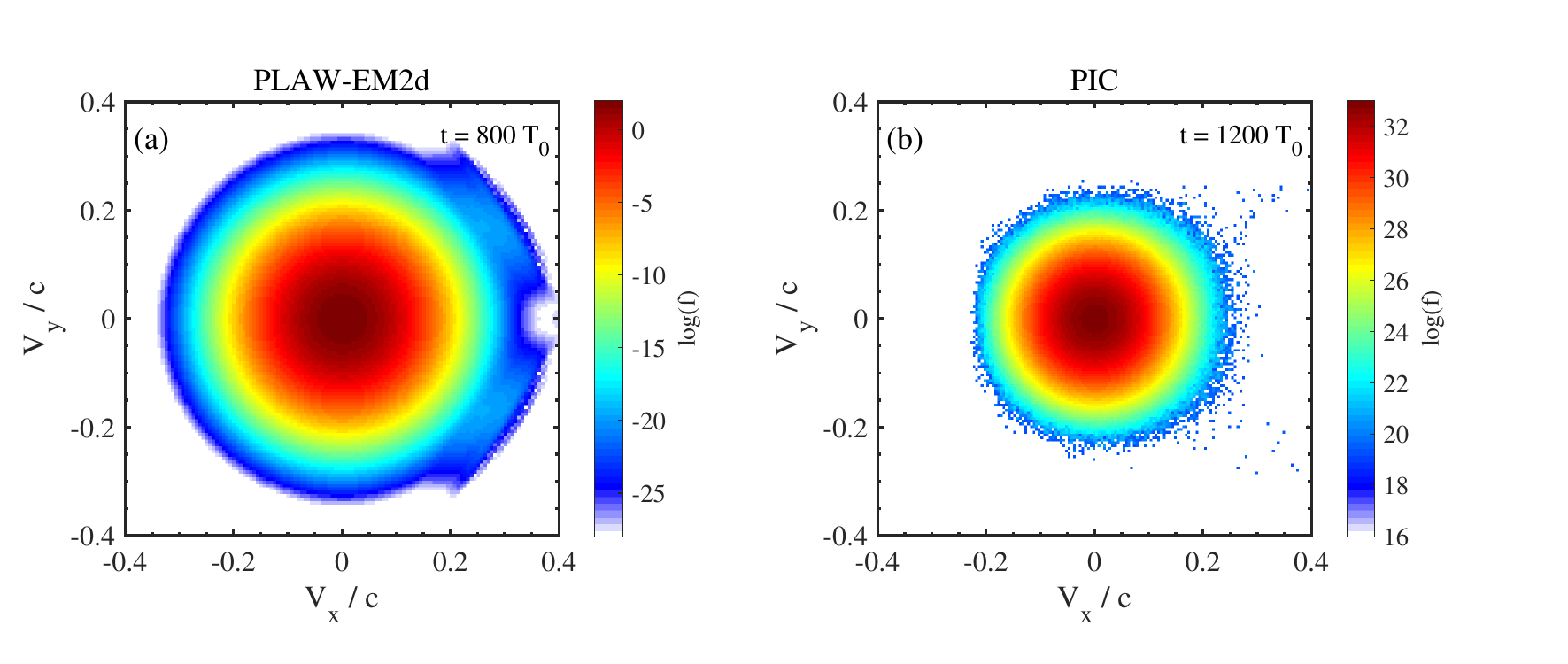}
\caption{\label{fig:side_vxvy} Reduced electron \(v_x\)-\(v_y\) distributions at \(t\approx800\,T_0\): \texttt{PLAW-EM2d} (left) and PIC (right). Both use \(|v_x|,|v_y|\le0.4c\) and show comparable broadening; the PIC result is noisier.}
\end{figure}

Figure~\ref{fig:side_vxvy} gives the reduced electron velocity distribution averaged over the central interaction region. At the present pump intensity of \(1.0\times10^{15}\ \mathrm{W/cm^2}\) (\(a_0 \approx 0.0095\)), both the Vlasov and PIC distributions remain centered near \((v_x,v_y) = (0,0)\). The interaction therefore does not produce a large bulk electron drift or strong ponderomotive channeling. Both panels nevertheless show broadening and distortion relative to the initial Maxwellian, which would be a circular Gaussian centered at the origin. The velocity support extends beyond the initial thermal width, consistent with electron heating and weak trapping by the oblique Langmuir wave. A slight asymmetry also appears along \(v_x\), with a tail toward positive \(v_x\), as expected from Landau resonance with this mode. The resonance condition is \(k_x v_x + k_y v_y = \omega_L\). Using \(\omega_L \approx 0.328\omega_0\), \(k_x = 0.9487\,k_0\), and \(k_y = 0.5933\,k_0\), the phase velocity along the Langmuir-wave direction is \(v_{\phi}=\omega_L/|\mathbf{k}_L| \approx 0.293c\). At \(v_y=0\), the resonance line intersects the \(v_x\) axis at \(v_{x,\mathrm{res}}=\omega_L/k_x \approx 0.35c\). Because Fig.~\ref{fig:side_vxvy} is plotted in the \((v_x,v_y)\) plane and the visible tail lies along positive \(v_x\), the latter value is the more relevant one for interpreting the horizontal distortion. Both velocities are well above \(v_{te} \approx 0.044c\), so the high-energy tail should respond most strongly. The weak distortion of the bulk remains consistent with the low pump intensity and with dynamics in the linear or weakly nonlinear regime.

As in the uniform-plasma benchmark, the Vlasov distribution is considerably smoother than the PIC result, which contains granular fluctuations from finite macroparticle sampling. On larger scales, however, the two distributions occupy similar velocity ranges and have comparable contour shapes. The central peak and broadened wings appear in both calculations. This common response supports the interpretation that the finite-\(|k_y|\) branches belong to a physical Raman side-scattering channel rather than arising from numerical artifacts or PIC noise.

\begin{figure}[htbp]
\centering
\includegraphics[width=0.68\textwidth]{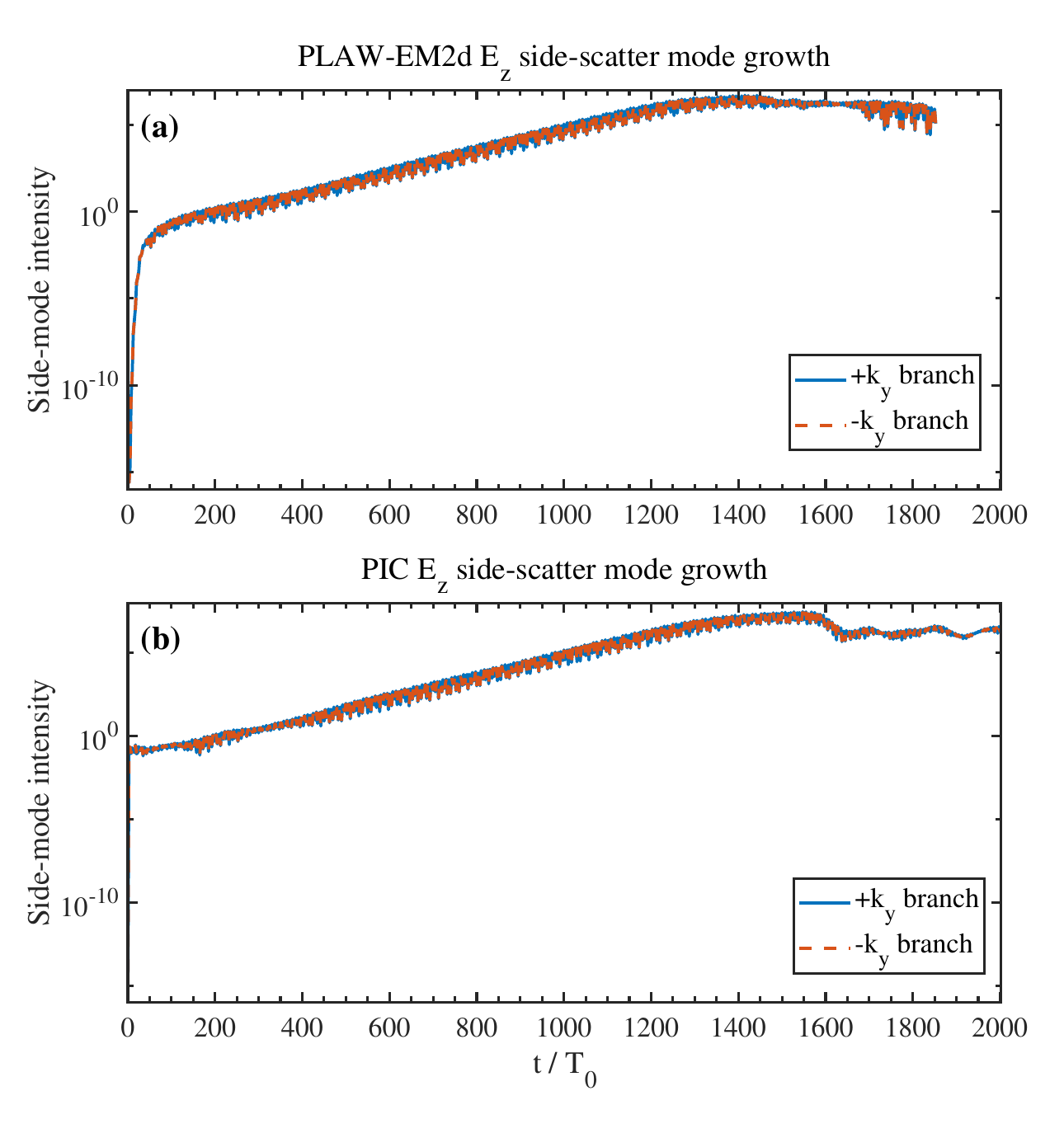}
\caption{\label{fig:side_mode} Time history of the integrated side-scatter spectral intensity extracted from the \(E_z\) spectrum. Top panel: \texttt{PLAW-EM2d}. Bottom panel: corresponding PIC result from the same side-scattering setup. In each panel, the solid curve tracks the \(k_y > 0\) branch near \((k_x,k_y) \approx (0, +0.5933)\,k_0\), and the dashed curve tracks the \(k_y < 0\) branch near \((k_x,k_y) \approx (0, -0.5933)\,k_0\). In both cases, the side-scatter signal turns on near \(t \approx 300\,T_0\) and then exhibits sustained growth over the main observation window; the Vlasov history is smoother, whereas the PIC history contains stronger fluctuations from finite-particle noise.}
\end{figure}

Figure~\ref{fig:side_mode} shows the time dependence of the side-scattered light to the snapshot diagnostics. It follows the integrated \(E_z\) spectral intensity in narrow windows around the predicted peaks \((k_x,k_y) \approx (0, \pm 0.5933)\,k_0\). In the Vlasov and PIC simulations, both branches are weak at early time and begin to rise near \(t \approx 300\,T_0\). The signals continue to increase through most of the recorded interval instead of appearing only as a short single-pass burst. The positive- and negative-\(k_y\) branches grow at similar rates and differ only moderately in amplitude. Particle noise makes the PIC curves less smooth, but their onset, overall growth, and saturation times agree with the Vlasov histories. For the present finite simulation box, this behavior is closer to an absolute-like side-scattering response than to a purely convective side-loss process. The diagnostic is an integrated spectral intensity, not a direct measure of energy. Figure~\ref{fig:side_mode} should therefore be used as evidence for the persistence and amplification of the modes rather than interpreted as an energy history.

The time history shows that the finite-\(|k_y|\) peaks belong to a sustained Raman side-scattering process rather than to isolated Fourier features or numerical noise. The \(+k_y\) and \(-k_y\) branches emerge at nearly the same time, consistent with emission in both transverse directions. This behavior follows the two-dimensional SRS matching geometry, in which the Langmuir wave has a finite transverse wave number and the scattered light propagates obliquely to the pump. The boundary conditions also affect the late-time response. Although the solver supports either open or periodic boundaries in \(y\), the present calculations use the periodic option to limit computational cost. Side-scattered light that leaves one transverse boundary can then re-enter through the other and undergo further amplification. This recirculation reinforces late-time growth and makes the response appear more absolute-like in the finite computational domain. A substantially wider transverse domain would reduce this effect but would also make the fully kinetic two-dimensional Vlasov calculation prohibitively expensive. The present setup is intended to resolve the self-consistent side-scattering physics at an affordable cost. A quantitative large-aperture calculation with weaker recirculation is left for future work.

\section{Conclusion}\label{conclu}

We have developed and tested \texttt{PLAW-EM2d}, a two-dimensional Vlasov-Maxwell code for kinetic studies of laser-plasma instabilities. The model couples an Eulerian Vlasov solver to a hybrid electrostatic-\(TE_z\) field decomposition. Vacuum propagation and Landau damping validate the separate electromagnetic and electrostatic branches, while the uniform-plasma PIC comparison tests their coupled kinetic response. In the density-gradient target, the code captures convective SRS backscatter, with a reflectivity that agrees in order of magnitude with the Rosenbluth estimate. In the transversely extended target, it resolves oblique electromagnetic sidebands and Langmuir peaks at the wave vectors required by the two-dimensional resonance conditions. To our knowledge, this is the first self-consistent demonstration of SRS side scattering in a two-dimensional Vlasov simulation. \texttt{PLAW-EM2d} can therefore be used for multi-dimensional kinetic studies of SRS. Extending it to stimulated Brillouin scattering and other ion-mediated instabilities would require an additional kinetic species and longer simulations. Larger, more realistic laser-plasma geometries will also require further algorithmic optimization and high-performance computing.

\section{Acknowledgements}

This work was supported by  National Natural Science Foundation of China (Grant Nos.12405277,U2530214 and 12475237), the Scientific Research Foundation for High-level Talents of Anhui University of Science and Technology (Grant No.2022yjrc106),  Anhui Provincial Natural Science Foundation (Grant No.2308085QA25), the Strategic Priority Research Program of Chinese Academy of Sciences (Grant No.XDA25050700) and the Fund of National Key Laboratory of Plasma Physics (Grant No.6142A04230103).


\end{document}